\documentclass[fleqn,usenatbib]{mnras}

\usepackage{newtxtext,newtxmath}
\usepackage[T1]{fontenc}

\DeclareRobustCommand{\VAN}[3]{#2}
\let\VANthebibliography\thebibliography
\def\thebibliography{\DeclareRobustCommand{\VAN}[3]{##3}\VANthebibliography}

\usepackage{graphicx}	% Including figure files
\usepackage{amsmath}	% Advanced maths commands
\newcommand{\rh}{R_{\mathrm{h}}}
\newcommand{\rhbh}{R_{\mathrm{h,BH}}}
\newcommand{\rc}{R_{\mathrm{c}}}
\newcommand{\ncbh}{N_{\mathrm{c,BH}}}
\newcommand{\frh}{f_{\mathrm{rh}}}
\newcommand{\fcbh}{f_{\mathrm{c,BH}}}
\newcommand{\fb}{f_{\mathrm{b,0}}}

\newcommand{\rhave}{\langle R_{\mathrm{h,B0}} \rangle}
\title[The impact of primordial binary]{The impact of primordial binary on the dynamical evolution of intermediate massive star clusters}

\author[Long Wang et al.]{Long Wang, $^{1,2}$\thanks{E-mail:long.wang@astron.s.u-tokyo.ac.jp}
  Ataru Tanikawa$^{3}$ and 
  Michiko S. Fujii$^{1}$
  \\
$^{1}$Department of Astronomy, School of Science, The University of Tokyo, 7-3-1 Hongo, Bunkyo-ku, Tokyo, 113-0033, Japan \\
$^{2}$RIKEN Center for Computational Science, 7-1-26 Minatojima-minami-machi, Chuo-ku, Kobe, Hyogo 650-0047, Japan\\
$^{3}$Department of Earth Science and Astronomy, College of Arts ans Sciences, The University of Tokyo, 3-8-1 Komaba, Meguro-ku, Tokyo 153-8902, Japan \\
}

\date{Accepted XXX. Received YYY; in original form ZZZ}

\pubyear{2015}

\begin{document}
\label{firstpage}
\pagerange{\pageref{firstpage}--\pageref{lastpage}}
\maketitle

% Abstract of the paper
\begin{abstract}
Observations found that star clusters contain a large fraction of binaries.
Tight binaries are an important heating source that influences the long-term dynamical evolution of star clusters. 
However, due to the limitation of $N$-body tool, previous theoretical modelling for globular clusters (GCs) by using direct $N$-body simulations have not investigated how a large fraction of primordial binaries affect their long-term evolution.
In this work, by using the high-performance $N$-body code, \textsc{petar}, we carry out star-by-star models for intermediate massive GCs ($N=100000$) with the primordial binary fraction varying from 0 to 1.
We find that when a stellar-mass black hole (BH) subsystem exists, the structural evolution of GCs (core and half-mass radii) only depends on the properties of massive primordial binaries, because they affect the number of BH binaries (BBHs), which dominate the binary heating process.
Low-mass binaries including double white dwarf binaries (BWDs) have almost no influence on the dynamics.
Meanwhile, %the dynamical perturbation also change the orbits of binaries and result in mergers. 
%The mergers of compact objects in dense star clusters like globular clusters (GCs) can generate gravitational waves (GWs).
only gravitational wave (GW) mergers from BBHs are strongly affected by dynamical interactions, while low-mass mergers from BWDs show no difference in the isolated environment (field) and in GCs.
Low-mass binaries become important only after most BHs escape and the core collapse of light stars occurs.
Our result suggests that for $N$-body modelling of GCs with a black hole subsystem dominating binary heating, it is not necessary to include low-mass binaries.
These binaries can be studied separately by using standalone binary stellar evolution codes. 
This way can significantly reduce the computing cost.
%Even the stochastic effect of initial condition (random seed) causes a larger scatter of the half-mass radius than that from a different $f$.

\end{abstract}

% Select between one and six entries from the list of approved keywords.
% Don't make up new ones.
\begin{keywords}
methods: numerical -- galaxies: star clusters: general -- stars: black holes
\end{keywords}

%%%%%%%%%%%%%%%%%%%%%%%%%%%%%%%%%%%%%%%%%%%%%%%%%%

%%%%%%%%%%%%%%%%% BODY OF PAPER %%%%%%%%%%%%%%%%%%

\section{Introduction}

% observation of binary formation in star-forming region
The observation shows that young star clusters contain a high fraction of multiplicity. \cite{Sana2012} found that more than 70 percent of O-stars in Galactic open stellar clusters have binary interaction.
\cite{Moe2017} studied the period and mass-ratio distribution of binaries in detail and found that at least 10 percent of stars are in hierarhcial multiple systems \citep[see also review from][]{Duchene2013}.
Thus, the multiplicity of stars is the major content in young star-forming stellar systems.

The property of multiplicity at the birth time is also important for understanding the star formation.  
Whether multiplicity shows an universal property is still an open question \citep[e.g.][]{Duchene2013}.
\cite{Duchene2018} showed that the frequency of stellar multiplicity in young star-forming regions (e.g. Orion nebula cluster) is systematically higher than that in field stars, but the reason is still unclear.
Meanwhile, %the binary fraction observed in globular clusters (GCs) is also significantly lower than that in young star-forming regions.
using high-resolution photometric observational data from the Hubble Space Telescope (HST), \cite{Sollima2007} found that the binary fractions are 10-50 percent for 13 low-density GCs; and \cite{Milone2012} found that nearly all binary fractions in a uniform photometric samples of 59 GCs are lower than that in the field.
\cite{Kroupa1995a,Kroupa1995b} assumed an universal property of primordial binaries in all kind of star-forming regions and found that the dynamical disruption of binaries in dense stellar system can explain the difference of binary fractions in star clusters and in the field.
Therefore, the stellar dynamics is important on reshaping the binary population.

Massive binaries also significantly affect the entire evolution of star clusters.
During the star-forming phase, the feedback from OB-stars (radiation, stellar winds and supernovae) can lead to the gas expulsion that quench the star formation \citep[e.g.,][]{Kroupa2001b, Goodwin2006, Baumgardt2007, Kroupa2018, Wang2019, Dinnbier2020, Semadeni2020, Fujii2021b} and further affect the large-scale molecular cloud \citep[e.g.,][]{Blaauw1964,Brown1994,Madsen2006,Odell1967,Odell2011,Ochsendorf2015,Kounkel2020,Grossschedl2021}.
When OB-stars are in binaries, few-body interactions can eject the feedback sources out of the birth place in a short timescale.
Thus, the formation of star clusters are very sensitive to runaway OB stars \citep[e.g.][]{Fujii2021b}. 
Since the feedback suddenly stops, the gas can continue to sink and to generate new stars. 
This may trigger multiple times of star formation as observed in the Orion nebular cluster \citep[][]{Kroupa2018,Wang2019}.
The few-body interactions can also lead to the mergers of binaries, which may also explain the formation of multiple stellar populations in GCs \citep[e.g.][]{deMink2009,Wang2020a}.
%In addition, the mergers of massive binaries may also explain the formation of multiple stellar populations in GCs 
%\cite{Torniamenti2021} also studied the impact of binaries on the early evolution of gas embedded young star clusters.

% impact of binary dynamics on long-term evolution
A part of massive stars finally evolve to black holes (BHs), which are much more massive than other stars.
These heavy population significantly influences the long-term dynamical evolution of star clusters \cite[e.g.][]{Spitzer1987,Binney1987,Breen2013,Kremer2018,Giersz2019,Antonini2020,Wang2020b}.
In the center of star clusters, the formation of binary BHs (BBHs) and the interactions between them and surrounding objects provide the dynamical heating that prevents the infinite core collapse and subsequently expands the whole star clusters.
These BBHs can be also the progenitors of the gravitational wave (GW) events \citep[e.g.][]{Ziosi2014,Kumamoto2019,DiCarlo2019,OLeary2009,Banerjee2020a,PortegiesZwart2000,Downing2010,Tanikawa2013,Bae2014,Rodriguez2016a,Rodriguez2016b,Fujii2017,Askar2017,Park2017,Samsing2018,Hong2020,Wang2021} observed by Advanced LIGO/VIRGO \citep{Abbott2019,Abbott2020}.

%Some numerical studies about the dynamical evolution of star clusters assume a 100 percent primordial binary fraction 
There are several theoretic studies about the impact of a high binary fraction up to 100 percent on the long-term evolution of star clusters.
\cite{Heggie2006} carried out direct $N$-body models for low-mass star clusters with no stellar evolution and an equal mass of stars. 
They showed that the different binary fractions do not have a strong impact on the global dynamical evolution after core collapse. 
%There are also a series of studies how a high binary fraction affect the evolution of GCs by using the fast Monte-Carlo method. %based on the model from \cite{Kroupa1995a,Kroupa1995b}.
By using the fast Monte-Carlo method, \cite{Ivanova2005} found that for a typical dense GC like 47 Tuc, only 5-10 percent is retained at the present-day due to the dynamical disruption of most wide binaries.
\cite{Hypki2013} and \cite{Belloni2019} found that the assumption of the property for primordial binaries from \cite{Kroupa1995a,Kroupa1995b} and \cite{Sana2012} can influence the formation of exotic objects such as blue stragglers and cataclysmic variables, respectively.
In addition, this assumption also results in a better fitting to the observed binary fraction inside and outside half-mass radii of GCs \citep{Leigh2015} and the color magnitude diagrams of GCs \citep{Belloni2017}.

However, the Monte-Carlo method approximates dynamical interactions. 
By comparing with direct $N$-body simulations, the Monte-Carlo method sometimes shows a different behaviour of core evolution \citep{Rodriguez2016c,Rodriguez2018}.
Therefore, it is necessary to carry out less approximated star-by-star $N$-body simulations to validate the results.
Due to the computational bottleneck of evolving binary orbits, such simulations for GCs with high binary fractions have not yet been done.
The previous largest $N$-body models for GCs \citep[Dragon models;][]{Wang2016} only contains a binary fraction of 5 percent. 
By using the recently released $N$-body code \textsc{petar} \citep{Wang2020d}, for the first time, we carry out a series of star-by-star $N$-body simulations for intermediate massive star clusters with different binary fractions up to 100~percent.

First, in Section~\ref{sec:method}, we describe the numerical $N$-body code, \textsc{petar}, and the initial condition for the star clusters.
Then, in Section~\ref{sec:result}, we analyze how binaries affect the structural evolution, binary heating and GW mergers in detail. 
Finally, we summarize and discuss our findings in Section~\ref{sec:conclusion}.
% GW mergers

\section{Methods}
\label{sec:method}
\subsection{\textsc{petar}}

In this work, we use the $N$-body code \textsc{petar} \citep{Wang2020c} to perform the numerical simulations of the star clusters.
\textsc{petar} is a high-performance $N$-body code that combines the particle-tree particle-particle method \citep{Oshino2011} and the slow-down algorithmic regularization method \citep[SDAR;][]{Wang2020b}.
The latter is the algorithm for accurately evolving the dynamical evolution of multiplicity. 
By using a hybrid parallel method based on the \textsc{fdps} framework \citep{Iwasawa2016,Iwasawa2020,Namekata2018}. 
The code can handle the simulation of massive star clusters with a binary fraction up to 100~percent.
This unique feature allows us to carry out this study.

We also use the single and binary stellar evolution packages, \textsc{sse} and \textsc{bse}, in the simulations \citep{Hurley2000,Hurley2002,Banerjee2020b}.
These population synthesis codes can follow the stellar wind mass loss, the stellar type changes and the mass transfer between interacting binaries.
The semi-empirical stellar wind prescriptions from \cite{Belczynski2010} is assumed.
We adopt the ``rapid'' supernova model for the remnant formation and material fallback from \cite{Fryer2012}, along with the pulsation pair-instability supernova \citep[PPSN;][]{Belczynski2016}.
Because supernova explosion is asymmetric, the compact objects forming after supernovae gain natal kick velocities.
Based on the measurement of the proper motions of field neutron stars (NSs) \citep{Hobbs2005}, we adopt a Maxwellian distribution of the kick velocity with a velocity dispersion of 265 km~s$^{-1}$.
The randomly sampling of this distribution result in high velocities that most exceed the escaping velocity of star clusters.
The fallback scenario assumes that a part of BHs do not have less or zero kick velocity due to mass fallback or failed supernovae, thus these BHs can be retained in star clusters. 
Meanwhile, electron-capture supernovae that result in NSs also have a low kick velocity (assuming a velocity dispersion of 3 km~s$^{-1}$), thus a part of NSs also can be retained.
The gravitational wave driven mergers of compact object binaries are also included in the \textsc{bse} following the approximation from \cite{Peters1964}.

\subsection{Initial conditions}

%To set up the realistic initial condition of star cluster is complicated, which needs to consider the gas component and the large-scale influence.
It is complicated to set up realistic initial conditions of star clusters including the gas component and its large-scale influence.
In this study, we focus on how the different binary fractions affect the long-term dynamical evolution of star clusters and GW events.
Thus, we ignore the complexity of the gas embedded phase and start the simulation in a gas-free and virialised initial condition which is commonly adopted in previous studies \citep[e.g.,][]{Wang2016}.

For all models in this work, the positions and velocities of stars and center-of-the-masses of binaries are generated by randomly sampling from the Plummer density profile \citep[][]{Plummer1911} with an initial half-mass radius of 2~pc.
The masses of stars are assigned by randomly sampling from the \cite{Kroupa2001} initial mass function (IMF) with the minimum and the maximum masses of $0.08$ and $150~M_\odot$, respectively.
To keep the shape of the IMF, we firstly sample the masses of all stars, and then, pair the two components in binaries based on the mass ratio distribution.
No tidal field is assumed.
We adopt a common metallicity of $Z=0.001$.
The modified \textsc{mcluster} code\footnote{https://github.com/lwang-astro/mcluster} is used to generate the initial conditions.

We have investigated 7 different initial fractions of primordial binaries ($\fb$), as listed in Table~\ref{tab:binit}.
The B0 set has no primordial binary; other sets except the B1.0 adopt the flat distribution of semi-major axes ($a$) and the thermal distribution of eccentricities ($e$).
The minimum and the maximum semi-major axes are $3~R_\odot$ and $100$~AU, respectively.
This range covers $a$ of the mass-transfer binaries and all tight binaries that can survive for a long term in the star clusters.
Due to the Heggie-Hills law \citep{Heggie1975,Hills1975}, wide (soft) binaries tend to be disrupted after a few strong encounters with intruders while tight (hard) binaries tend to be tighter after encounters.
The maximum $a$ of hard binaries can be estimated as
\begin{equation}
  a_{\mathrm{h}} = \frac{G m_1 m_2}{\langle m \rangle \sigma^2} 
  \label{eq:hhlaw}
\end{equation}
where $G$, $m_1$, $m_2$, $\langle m \rangle$ and $\sigma$ are gravitational constant, masses of two binary components, the local average mass and the local velocity dispersion, respectively.
In our model, the average $a_{\mathrm{h}}\approx32$~AU. 
Thus, a part of primordial binaries are in the wide region.

The random sampling from the distributions of $a$ and $e$ can form binaries where two components touch each other at the peri-center position. 
Such binaries should already have mass transfer or merged during the star formation and should not be included as the primordial binaries.
We apply the eigenevolution method suggested by \cite{Kroupa1995b} to adjust the orbital parameters of these binaries to avoid the forbidden region of $a$-$e$.

The pairing of masses of the two components in binaries is different for massive OB ($>5 M_\odot$) and low-mass stars.
We adopt the uniform mass ratio distribution between 0.1 and 1.0 for massive binaries following \cite{Sana2012} and the random pairing of masses for the low-mass binaries.
In addition, the observation shows a high multiplicity frequency of OB-stars \citep{Moe2017}, thus we adopt 100 percent fraction of OB binaries which is independent of $\fb$.

The B1.0 set follows the properties of primordial binaries of low-mass stars ($<5 M_\odot$) derived by \cite{Kroupa1995a,Kroupa1995b} and \cite{Belloni2017} and of high-mass stars based on \cite{Sana2012}.
\cite{Belloni2017} updated the model in order to be consistent with the observed color distribution of GCs.

\begin{figure}
    \centering
    \includegraphics[width=0.9\columnwidth]{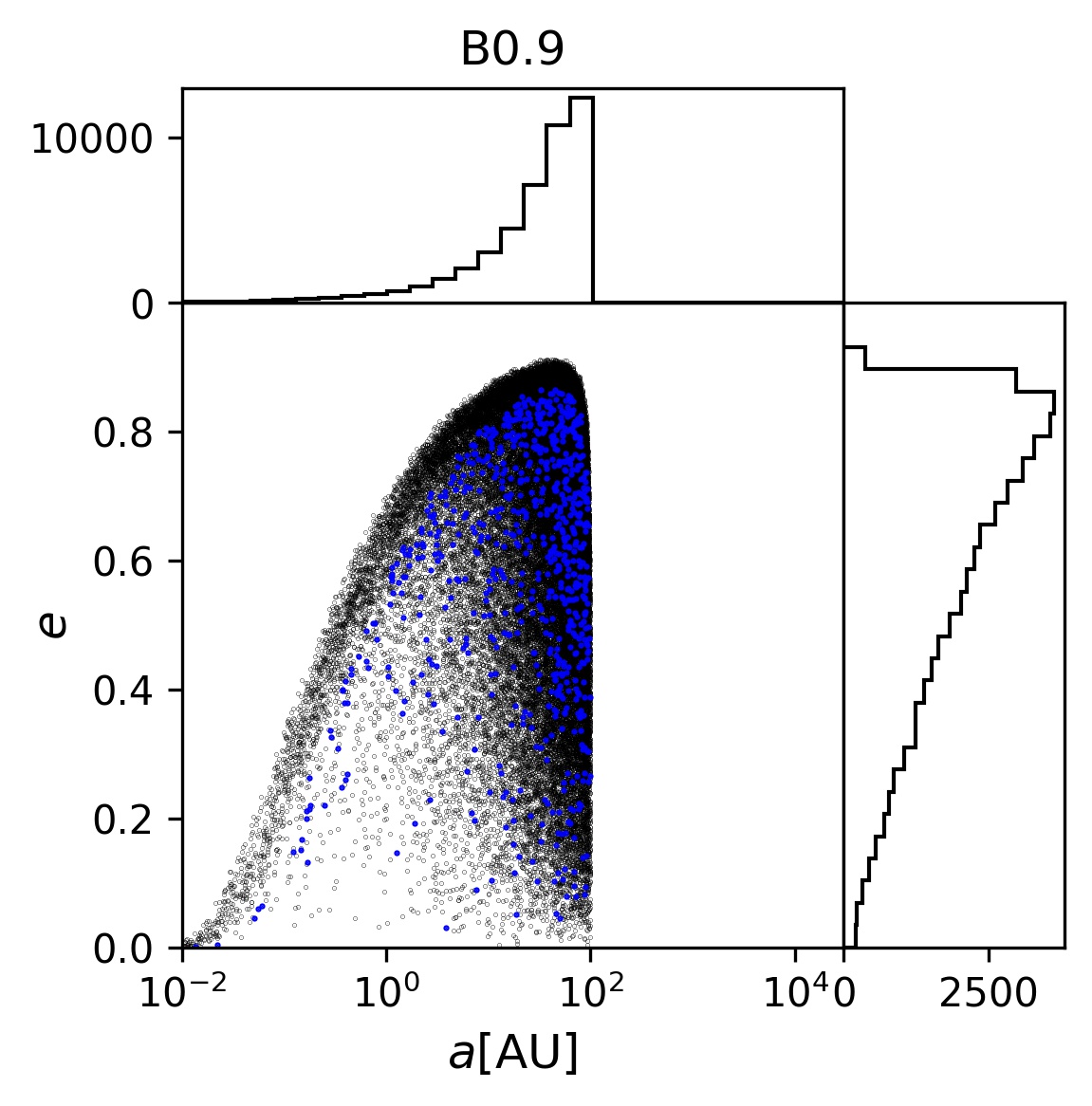}\\
    \includegraphics[width=0.9\columnwidth]{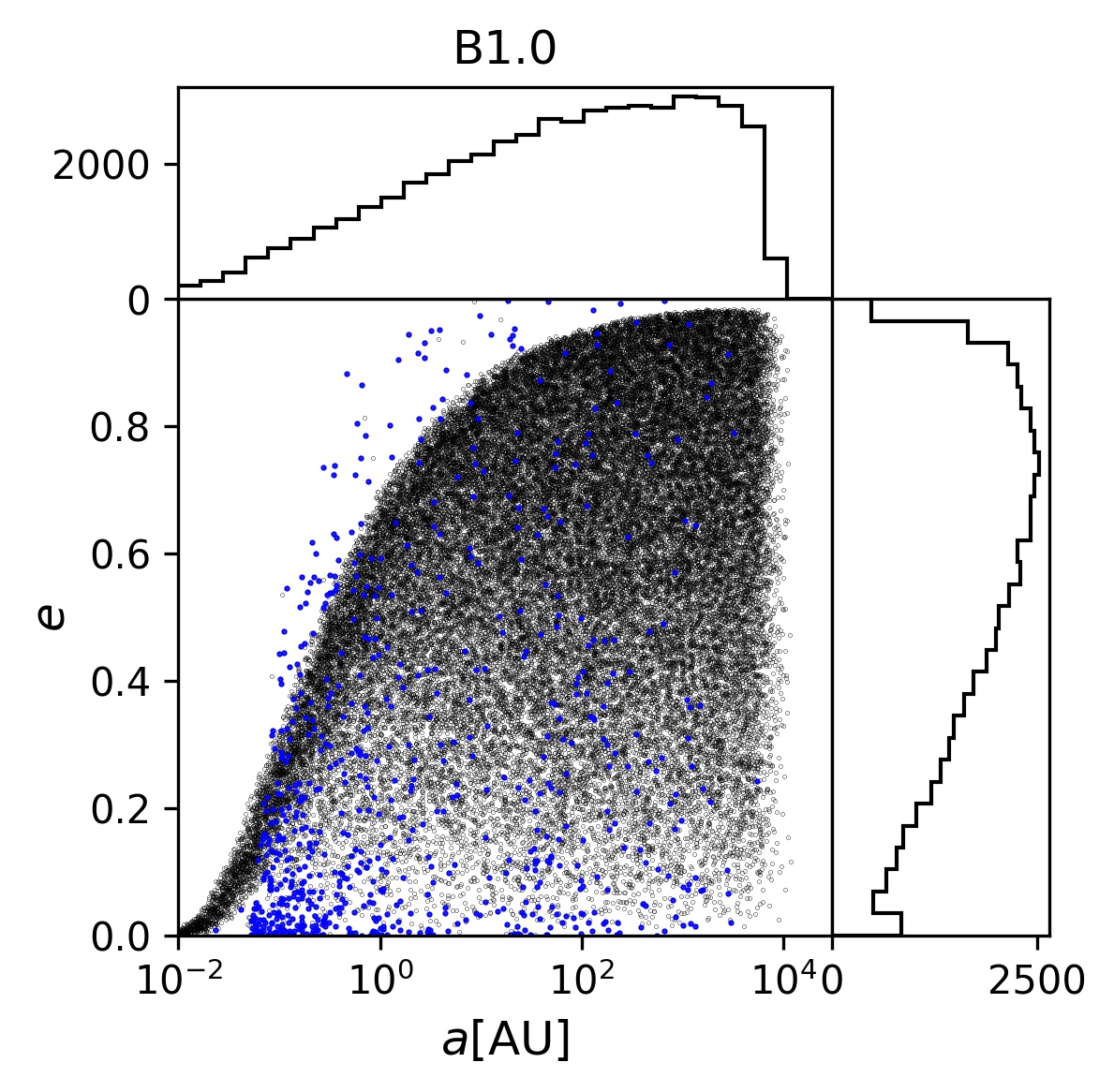}
    \caption{The $a$-$e$ distribution of the B0.9 (upper) and the B1.0 (lower) sets. In each panel, the middle plot shows the $a$-$e$ of all binaries; the upper and the right plots show the histograms of $a$ and $e$, respectively. Note that the ranges of $x$-axes of the two panels are different. 
    The black and blue dots represent the low-mass and massive binaries, respectively.}
    \label{fig:binit}
\end{figure}

Figure~\ref{fig:binit} compares the $a$-$e$ distribution for the B0.9 set (also representing the sets with $0<\fb<1$) and the B1.0 set.
For the B0.9 set, most of binaries have $a>10$~AU with the $e$ peaked at 0.8.
The B1.0 set has a much larger region of $a$ with the maximum value being approximately $1\times10^4$~AU. 
The distribution of $e$ is flatter between 0.6 and 0.9 and has a clump near zero.

\begin{table}
    \centering
    \caption{The sets of initial fractions of primordial binaries for the $N$-body models. The B1.0 set follows the initial properties of binaries described in \citep{Kroupa1995a,Kroupa1995b,Sana2012,Belloni2017}. Other models except the B0 adopt the flat distribution of semi-major axes and the thermal distribution of eccentricities.}
    \begin{tabular}{cccccccc}
    \hline
        Name &  B0 & B0.1 & B0.3 & B0.5 & B0.7 & B0.9 & B1.0 \\
    \hline
        $\fb$ & 0 & 0.1 & 0.3 & 0.5 & 0.7 & 0.9 & 1.0*\\
    \hline
    \end{tabular}
    \label{tab:binit}
\end{table}

We create a group of models for each set listed in Table~\ref{tab:binit} to investigate the long-term dynamical evolution.
The group is named as "L-group".
The names of models combines the set names in Table~\ref{tab:binit} and the suffix "L", such as B0L.
Simulations of the B0L, B0.1L,  B0.3L and B0.5L models have reached 12 Gyr, while others have reached 7 Gyr.
After 7~Gyr, most of BHs have escaped from the systems, thus it is sufficient for the purpose of this research. 

The random sampling of the initial masses, positions and velocities of stars introduces a stochastic effect.
To investigate how this affects the long-term evolution, we create the second "S" group.
In this group, we create 5 models with different random seeds for each $\fb$.
The suffixes "S0, "S1", ..., "S5" are used in the names to distinguish the models.
These models have been evolved to 500 Myr.

In the L- and S-groups, the binary fractions and the mass pairing of massive and low-mass binaries are different.
For a more general study, we also create a group of models without the special treatment of massive binaries, i.e., the property of massive binaries follow that of the low-mass binaries.
This group is named as (randomly pairing) "R-group".
Similar to the S-group, the R-group contains 5 models with different random seeds for each $\fb$, and each model has beend evolved to 500 Myr. 
Table~\ref{tab:groups} summarizes the difference of the three groups.
\begin{table}
    \centering
    \caption{The difference of the three groups of models. From the left to the right: labels of groups, final times of the simulation ( $T_{\mathrm{f}}$), numbers of models for each $\fb$ ($N_{\mathrm{mod}}$), binary fractions for OB stars ($\fb$[OB]) and pairing methods for the masses of two components of OB binaries.}
    \begin{tabular}{lllll}
    \hline
        Groups & $T_{\mathrm{f}}$ & $N_{\mathrm{mod}}$ & $\fb$[OB] & mass [OB]\\
        \hline
          L    & 7-12 Gyr & 1 & 1 & Sana et al. (2012) \\
          S    & 500 Myr  & 5 & 1 & Sana et al. (2012) \\
          R    & 500 Myr  & 5 & same as $\fb$ & random pairing \\
    \hline
    \end{tabular}
    \label{tab:groups}
\end{table}
Hereafter, when we refer to all the models with the same $\fb$, we simply call them a model name without a suffix, such as B0, or B0.1.

\section{Results}
\label{sec:result}
\subsection{The structural evolution}
\label{sec:r}

\begin{figure}
    \centering
    \includegraphics[width=\columnwidth]{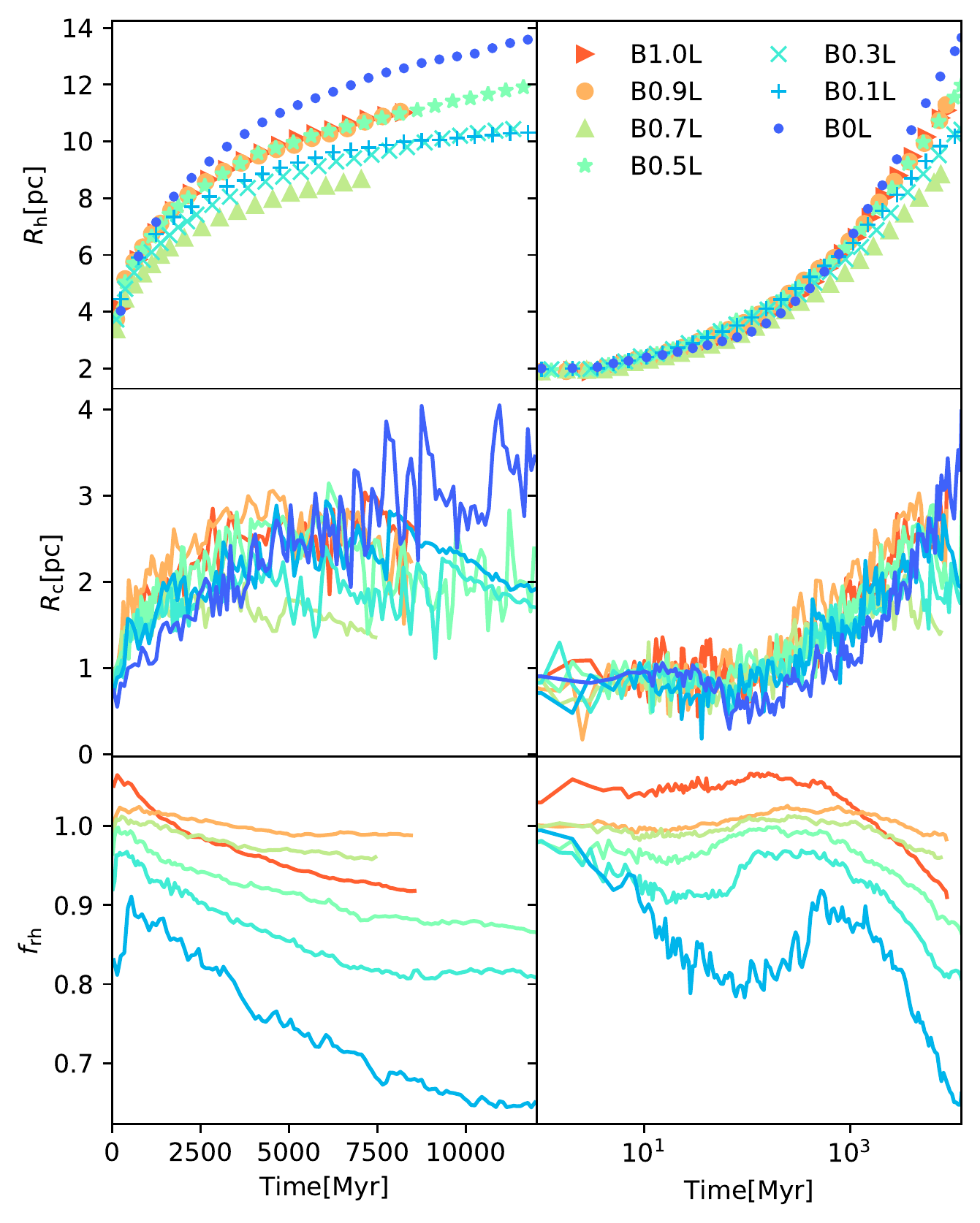}
    \caption{The comparison for the structural evolution for the L-group models. From the top to the bottom panels are the half-mass radii ($\rh$), core radii ($\rc$) and the ratio of half-mass radii between binaries and all stars ($\frh$). In the left and right panels, the horizontal axes are in linear and logarithmic scales, respectively.}
    \label{fig:rhrc}
\end{figure}

The principle of \cite{Henon1971,Henon1975} showed that in a gravitationally bound stellar system, the energy flux demanded by the global system should be balanced by the energy generated by the central engine (binary heating).
This principle links the dynamics of central binaries to the long-term evolution of the global structure.
\cite{Breen2013} showed that when the BH subsystem exists, BBHs are the major source of the binary heating.
The balance of the energy in the virial equilibrium can be reflected by the half-mass radii ($\rh$) of the global system and of the BH subsystem \citep[see Eq.~3 in ][]{Breen2013}:
\begin{equation}
    \frac{E}{E_{\mathrm{BH}}} \approx \frac{M^2 \rhbh}{ M_{\mathrm{BH}}^2 \rh}
\end{equation}
where the suffix "BH" reflects the properties of the BH subsystems.
When the BH subsystem exists, the core radius $\rc$ is linked to $\rhbh$.
Thus, in Figure~\ref{fig:rhrc}, we compare the evolution of $\rh$ and $\rc$ for all models in the L-group and investigate that how $\fb$ affects the structural evolution.

A scatter of $\rh$ among the models appears after around 100 Myr. 
However, there is no monotonic relation between $\rh$ and $\fb$.
The B0L and the B0.7L models have the largest and the smallest $\rh$, respectively,
while the B0.3L, B0.5L and B0.9L models have a similar $\rh$.
Thus, such a difference is not likely to be caused by different $\fb$.

%Considering the Henon's principle, the binary heating is controlled by the global energy requirement.
%Since the initial global structure is the same for all models and only the number of objects (binaries are not resolved) differs, the energy requirement is similar. 
%Thus, no matter what $\fb$ is, only the massive binaries (BBHs) at the center contribute to the binary heating until they are ejected from the center, and the total energy provided by these binaries is fixed by the energy requirement.
%Because the L-group models except the B0 have the identical property of the massive primordial binaries, it is reasonable that the structural evolution is independent of $\fb$ for these models. 

We can understand the reason why the $\rh$ evolution is independent of $\fb$ as follows. 
$\rh$ expands because of binary heating at the cluster center. 
The binary heating is controlled by the global energy requirement, according to Henon's principle. 
The global energy requirement is similar among the models in Figure \ref{fig:rhrc}. 
This is because the initial global structure is the same for all the models, although only the number of objects differs. 
Note that a (tight) binary star is counted as one object, since it is not resolved in dynamics unless its orbit is significantly perturbed by a close encounter.
The global energy requirement is met by only the massive primordial binaries (here, BBHs and their progenitors). 
The massive primordial binaries have identical properties among the L-group models except for the B0 regardless of $\fb$. 
Thus, their $\rh$ evolution is independent of $\fb$.

%Both $\rh$ and $\rc$ show the expansion after 100 Myr.
%\cite{Breen2013} shows that BHs in star clusters form a dense sub-system in the %center after mass segregation.
%Then, BBHs dominate the binary heating process.
%As a result, the global system expands (see also \cite{Wang2020}). 

Binaries are relatively heavier than single stars.
Thus, binaries become more centrally concentrated in the cluster due to the mass segregation. 
The bottom panel of Figure~\ref{fig:rhrc} shows the ratio of the half-mass radii between binaries and all stars ($\frh$).
The value of 1.0 indicates that binaries have the same density profile as singles, while the value below 1.0 indicates the mass segregation of binaries.
The B1.0L model has $\fb>1.0$ initially due to the low-number statistic of single stars.
There is a trend that the models with higher $\fb$ show larger $\frh$ (smaller degree of mass segregation),
because the mass segregation firstly appears in the sub-group of massive binaries, and the larger fraction of low-mass binaries in these models smooths out the feature of mass segregation for all binaries. 

%\michiko{(MF: What is the initial rise of $\frh$ seen in all models?)}
There is an rise of $\frh$ around 100~Myr for all models, while models with lower $\fb$ show a more pronounced feature.
There are two mechanisms contributing to this rise. (1) the natal kicks of compact objects formed after supernovae of massive stars invert the mass segregation; these kicks mostly occurred before 100~Myr in massive binaries; (2) the binary heating from BBHs after core collapse ejects BHs from the center.

The evolution of $\rc$ is roughly identical for all models except the B0L.
Comparing the B0L and the other models, they show two major differences:  

(1) The B0L model shows a pronounced feature of core collapse at approximately 50 Myr (see the middle right panel of Figure~\ref{fig:rhrc}) while the others do not.
The binary heating is the major mechanism to suppress core collapse.
In the B0L model, such the process starts when the first binary forms after the core collapse.
In the other models, the primordial binaries can provide the heating.
The mass segregation starts to appear around 10~Myr as shown in the evolution of $\frh$ (see the bottom right panel of Figure~\ref{fig:rhrc}).
It is expected that the binary heating of massive binaries can already start at this time, and thus, no clear sign of core collapse appears around 50~Myr. 
%\cite{Heggie2006} found that the core collapse timescale is different for the models with and without primordial binaries but the structural properties are independent of $\fb$.
%Our result shows the consistent result. 

\begin{figure}
    \centering
    \includegraphics[width=\columnwidth]{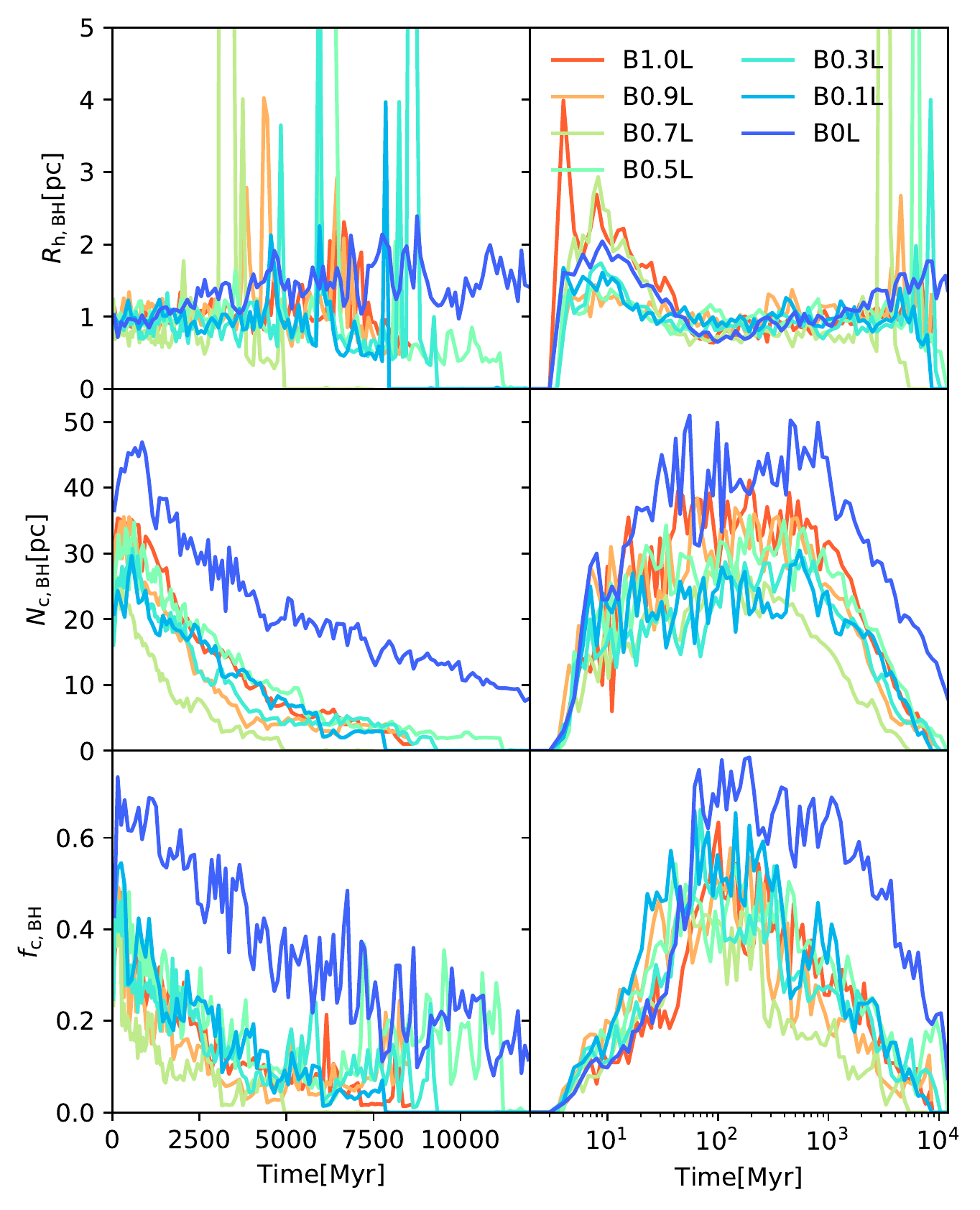}
    \caption{The comparison for the structural evolution of the BH subsystems for the L-group models. From the top to the bottom panels are half-mass radii of BHs ($\rhbh$), numbers of BHs inside $\rc$ ($\ncbh$) and the mass ratio of BHs and all objects inside $\rc$ ($\fcbh$).}
    \label{fig:bhrhrc}
\end{figure}

(2) After about 7~Gyr, $\rc$ of the B0L model continues increasing while those of the other models (B0.1L, B0.3L and B0.5L) show the signal of core collapse. 
To explain the reason, we show the properties of the BH subsystems in Figure~\ref{fig:bhrhrc}. 
The upper panels compare the evolution of the half-mass radius of BHs ($\rhbh$). 
For the B0L model, $\rhbh$ continues increasing till 12~Gyr, while for all the other models, $\rhbh$ decrease with sharp peaks appearing between 4 to 8~Gyr.
The evolution of the number of BHs inside $\rc$ ($\ncbh$; the middle panels of Figure~\ref{fig:bhrhrc}) explains the reason for this behaviour.
After about 100~Myr, the B0L model contains much more BHs than the other models.
At the end of simulation, the B0L model still keeps about 10 BHs inside $\rc$.
But BHs are already depleted between 4 to 8 Gyr in the other models.
Thus, the sharp peaks appear in the evolution of $\rhbh$ during this period because of the ejection of BHs and the low-number statistics.
After BHs are depleted, the leak of binary heating drives the core collapse of light stars as shown in Figure~\ref{fig:rhrc}. 
This is also consistent with the theory of \cite{Breen2013}.
Therefore, the initial difference of $\ncbh$ significantly affects the long-term structural evolution of star clusters.

%Since all OB stars are in binaries for the models with primordial binaries, many of the tight binaries have merged or escaped from the cluster via strong few-body interactions, and thus, after 100~Myr the number of BHs is much smaller compared to that of the B0L model. 
%Such initial difference significantly affects the long-term evolution of the BH subsystems. 
%Due to a initially low $\ncbh$ in the other models 
%$\ncbh$ decreases to 0 in 4-8 Gyr for all models with primordial binaries.

The bottom panel of Figure~\ref{fig:bhrhrc} shows the evolution for the mass ratio of BHs and all objects inside $\rc$ ($\fcbh$). 
After the BH subsystem forms, BHs contribute to about 40 and 60 percent of the mass inside $\rc$.
$\fcbh$ decreases as BHs escape via few-body interactions. 
At the end, BHs in the B0L model still occupy about 10 percent of the core mass. 

To further investigate the reason for the initially different number of BHs, we collect all BH progenitors with the initial stellar masses $>20.3~M_\odot$ and check how many of them have escaped from the star clusters before 100~Myr and how many have merged in binaries.
Such events reduce the remaining number of BHs in the cluster.
The result is shown in Table~\ref{tab:obesc}.
The numbers of formed BHs ($N_{\mathrm{BH}}$) differ among the models due to the random effect of initial condition and the mergers.
The major mechanism to remove BHs are from single escapers (E-S), where most of them are driven by the high natal kick velocities after asymmetric supernovae (E-SK).
The numbers of E-SK are similar for all L-group models, while the numbers of E-S and binary escapers (E-B) from the B0L model are about 10 less than those of the other models, respectively.
Meanwhile, there is no natal kick driven binary escapers (E-BK) for all models.
There are also a few less mergers in the B0L model.
Therefore, the escaping of singles and binaries driven by dynamical interactions and mergers are the major mechanism that causes the less number of BHs in the models with primordial binaries.

%In the next section, we collect all binaries of compact objects inside $\rc$ and discuss their contribution to the energy heating in detail.

\begin{table}
    \centering
    \caption{The number of merged and escaped massive stars with initial masses $>20.3~M_\odot$ for the L-group models. For column labels, $N_{\mathrm{BH}}$ indicates the total number of formed BHs; "Merge" indicates the binary mergers with both components being massive stars; "E-S" and "E-B" indicate the single and the binary escapers, respectively; "E-SK" and "E-BK" indicate the single and the binary escapers are caused by the natal kick after supernovae, respectively. For a binary escaper, if both components are massive stars, it counts twice.}
    \begin{tabular}{lllllll}
\hline
Model & $N_{\mathrm{BH}}$ & Merge &   E-S &  E-SK &   E-B &  E-BK \\
\hline
  B0L &   203 &     2 &    76 &    75 &     1 &     0 \\
B0.1L &   193 &     6 &    92 &    84 &    13 &     0 \\
B0.3L &   172 &     5 &    86 &    82 &    16 &     0 \\
B0.5L &   193 &     8 &    95 &    81 &    19 &     0 \\
B0.7L &   171 &     8 &    86 &    80 &    17 &     0 \\
B0.9L &   186 &     9 &    76 &    68 &    14 &     0 \\
B1.0L &   190 &     7 &    93 &    80 &    12 &     0 \\
\hline
    \end{tabular}
    \label{tab:obesc}
\end{table}

\subsection{Binary heating}

\begin{figure}
    \centering
    \includegraphics[width=\columnwidth]{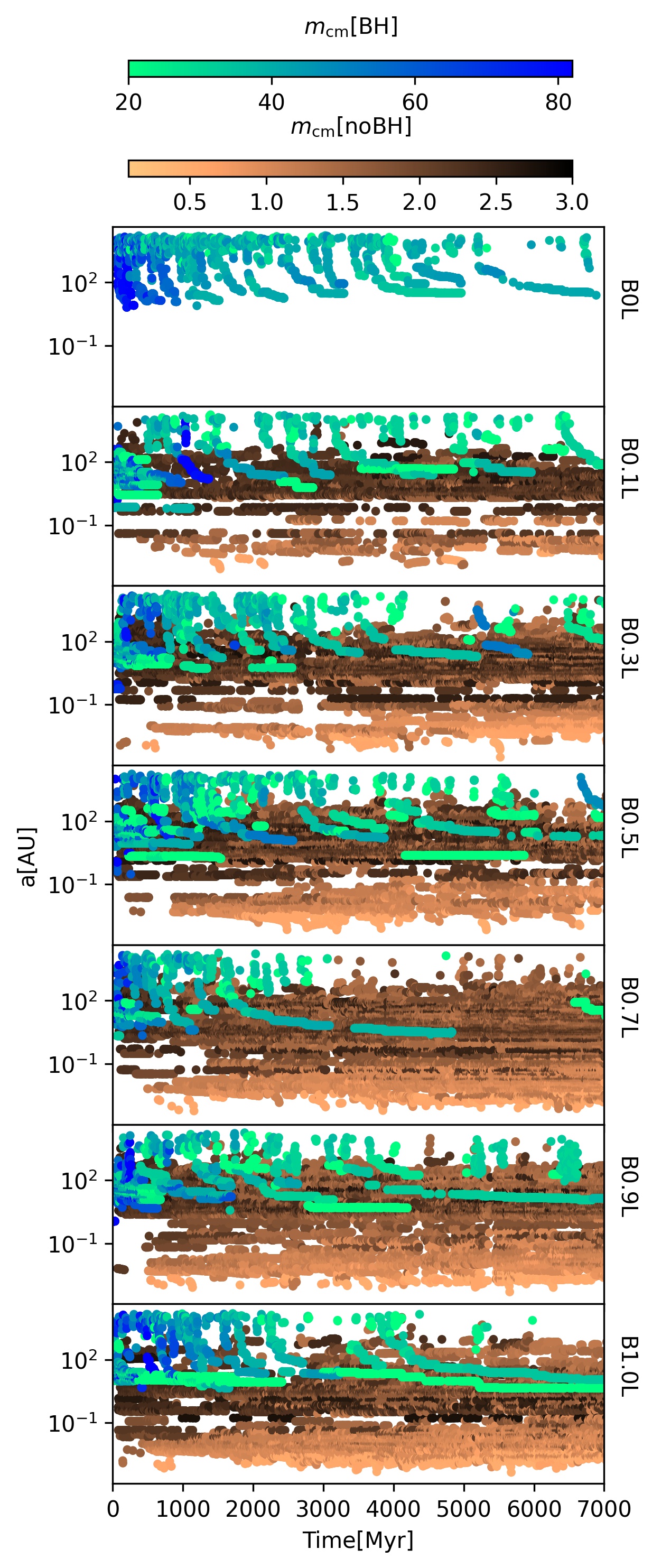}
    \caption{The evolution of semi-major axes ($a$) of all compact binaries inside $\rc$ for the L-group models. Each point represents a binary. Colors represent the masses of binaries. For a better view, binaries with and without BH (noBH) are shown in two types of color maps. The neighbor points with the same color can be identified as the evolutionary track of an individual binary.}
    \label{fig:cbinrc}
\end{figure}

%In the above section, we explain that the identical evolution of $\rh$ and $\rc$ for the L-group models (excluding the B0L) in Figure~\ref{fig:rhrc} is due to a similar heating from the massive binaries (BBHs). 
%To validate this hypothesis, we 
%The result is shown in Figure~\ref{fig:cbinrc}.
To validate that the massive binaries (BBHs) dominate the binary heating, we investigate the evolution of semi-major axes ($a$) of all compact binaries inside $\rc$ for the L-group models (see Figure~\ref{fig:cbinrc}).
Firstly, the tracks of $a$ for individual BBHs show a pronounced decrease of $a$ down to 1 -- 10 AU.
This is the sign of binary heating process as described by the the Heggie-Hills law, where tight binaries become tighter after encounters with others and transfer the binding energy outwards.
When $a$ becomes small enough, the BBH either merge or escape from the core. 
Generally, only a few BBHs can provide heating simultaneously. 
Once tight binaries in the core are ejected, new tight binaries start to evolve.

The B0L model continues producing new BBHs up to 7~Gyr, while the formation rate of tight BBHs significantly decreases in other models after 4~Gyr due to the lack of BHs. 
The total number of dynamically formed BBHs inside $\rc$ before 7~Gyr is about 570, where only 62 BBHs are tight with the minimum semi-major axis below 200~AU.
The BBHs formed via the exchange of members after three- or four-body encounters are also counted as new ones.

%New tight BBHs form after old ones disappear.
%For all models, when most OB stars evolve to compact objects (after around 60 Myr), BHs become the most massive objects.
%Thus, the binary heating is dominated by BBHs \citep{Breen2013}.
%In addition, 
%A long-exist tight BBH seems to suppress the formation of new tight BBH from 3 to 5 Gyr in the B0.7L model.

Meanwhile, there are also a large group of low-mass binaries with white dwarfs (WDs) and NSs in all models except the B0L.
They form from the low-mass primordial binaries.
%A NS-WD binary shows a darker color.
Unlike the BBHs, these low-mass binaries do not show the decrease of $a$, and thus, they do not contribute to the binary heating. 
A few exceptions with small $a$ are probably due to the binary stellar evolution, which leads to the mergers of WD-WD binaries at the end.
%This explains that the low-mass primordial binaries, it suggests that the l
This explains why the evolution of $\rh$ and $\rc$ is independent of $\fb$.

Figure~\ref{fig:cbinrc} only shows the result of the first 7~Gyr, when BBHs dominate the binary heating. 
After most BHs escape, the core collapse of light objects including NSs and WDs occur as shown in Figure~\ref{fig:rhrc}.
Then, low-mass binaries start to dominate the binary heating.

\subsection{The impact from massive binaries}

\begin{figure}
    \centering
    \includegraphics[width=\columnwidth]{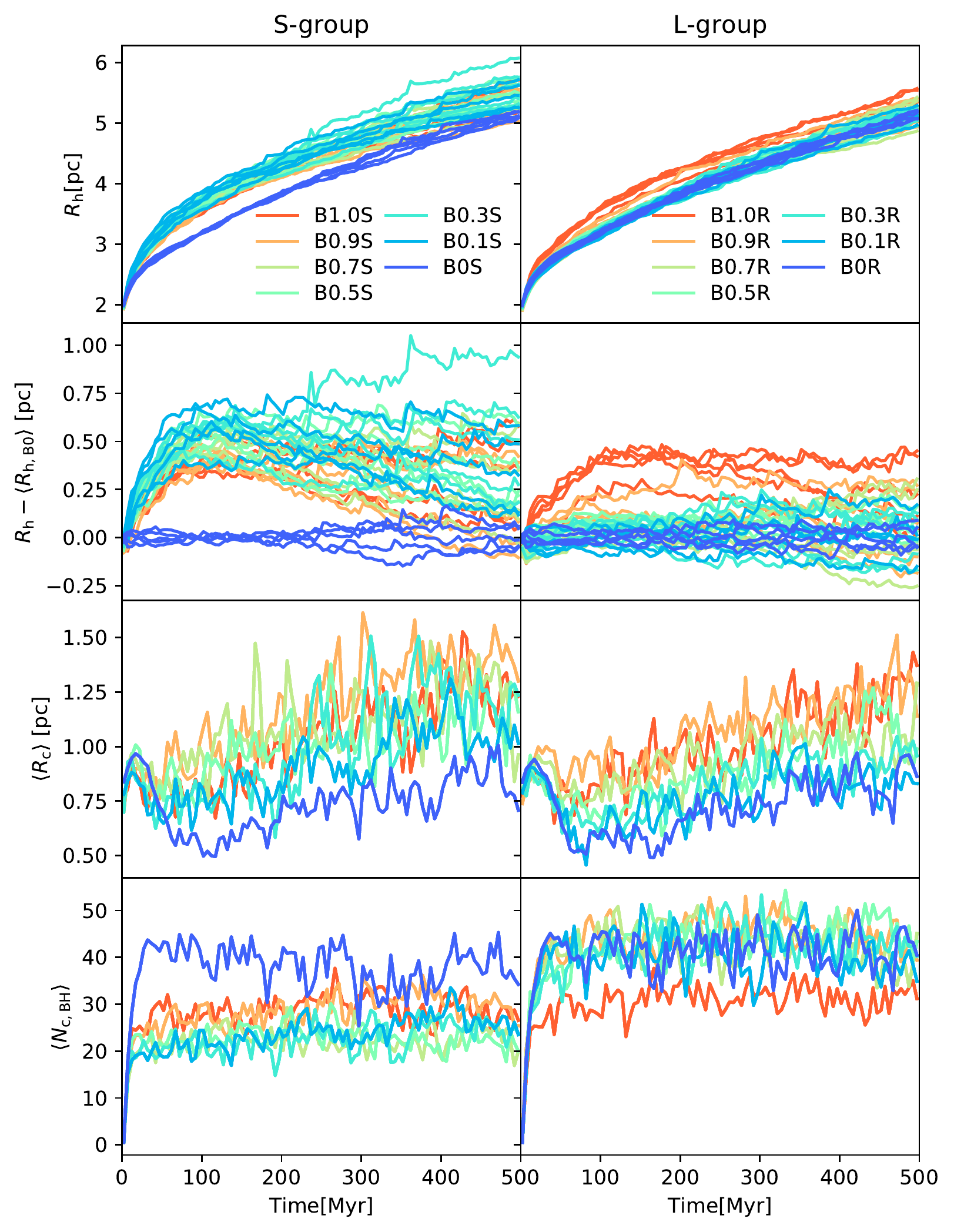}
    \caption{The comparison between the S- (left panels) and the R-groups (right panels) for the evolution of $\rh$,  $\rc$  and $\ncbh$ for all models. To indicate the scatter due to the stochastic effect, the evolution of $\rh$ from the 5 models for each $\fb$ are shown separately with the same color. To amplify the difference depending on $\fb$, for $\rc$ and $\ncbh$, the average values from the 5 models of each $\fb$ are shown instead. In addition, the second row of panels show the difference of $\rh$ between each model and the averaged $\rh$ of 5 B0 models ($\rhave$).}
    \label{fig:comps}
\end{figure}

As shown in Figure~\ref{fig:rhrc}, the structural evolution in the L-group does not monotonically depends on $\fb$.
However, it can be possible that the random sampling of masses, positions and velocities of individual stars and binaries cause the stochastic scatter, which may override the general $\fb$-dependent trend.
We investigate this by comparing the structural evolution of the S-group models in the left two panels of Figure~\ref{fig:comps}.
By showing the 5 models with different random seed for each $\fb$ together (same color), we can identify that the evolution of $\rh$ for the models with primordial binaries indeed has a wide stochastic scatter due to the random sampling (see the first and the second rows in Figure~\ref{fig:comps}).
Including such a scatter, the models with primordial binaries show an identical general trend, while the B0S models show a pronounced difference.
This again suggests that only the massive binaries (BBHs) affect the long-term structural evolution.

To further confirm this, we also compare the structural evolution of the R-group models, where the primordial binary fraction follows $\fb$, in the right panels of Figure~\ref{fig:comps}.
Unlike those in the S-group, the models excluding the B1.0R show a closer evolution of $\rh$ and $\rc$ to the B0R models, %The B1.0R model which used the \cite{Sana2012} 
The low-$\fb$ models also show a similar timescale of core collapse to that of the B0R model.

Figure~\ref{fig:bhrhrc} suggests that $\ncbh$ is the key parameter to determine the structural evolution. 
We also compare this for the S- and R-group models in the bottom panels of Figure~\ref{fig:comps}.
Similar to the L-group, the S-group shows an initially large difference of $\ncbh$ between the B0S models and the others.
But the R-group has a different behavior: except the B1.0R models, the others show a similar $\ncbh$ that weakly depends on $\fb$.
This is consistent with the behaviour of $\rh$ and $\rc$.
It is interesting to see that the large-$\fb$ models in the R-group do not show a pronounced difference of $\ncbh$ compared to that of the B0R models.
This suggests that the fraction of massive primordial binaries is not the major parameter that determines $\ncbh$, the orbital properties, including the mass ratio, the period distribution and the eccentricity distribution, have a stronger impact.
In summary, we can conclude that only massive binaries are important on the dynamics evolution of the system.
%But the difference in the S-group models except the B1.0R model is much smaller.

%This is expected, since the ejected numbers of BHs are similar (see the discussion about the Table~\ref{tab:obesc} in Section~\ref{sec:r}).
%The behaviours among $\ncbh$, $\rh$ and $\rc$ are self-consistent. 

\subsection{GW mergers}

\begin{table}
    \centering
    \caption{The merger numbers of different types for the L-group models. The column labels BWD, BNS, BBH, BWN, BWB and BNB indicate WD-WD, NS-NS, BH-BH, WD-NS, WD-BH and NS-BH mergers, respectively. For each model, we show a few sub-groups of mergers: ISO represents mergers via isolated binary stellar evolution by using the standalone \textsc{bse} code to evolve the primordial binaries from the initial condition; the prefixes "I" and "E" represent the mergers inside star clusters and the escaped ones, respectively; the suffix "C" indicates that the mergers occurred in both ISO and $N$-body models; and the suffixes "PB" and "DY" represents the progenitors are primordial binaries and dynamically formed binaries, respectively. }
    \begin{tabular}{llcccccc}
\hline
Model &     Type &   BWD &   BNS &   BBH &  BWN &  BWB &  BNB \\
\hline
      &     I-DY &     0 &     0 &     2 &     0 &     0 &     0 \\
      &     E-DY &     0 &     0 &     0 &     0 &     0 &     0 \\
\hline
      &      ISO &    10 &     0 &     0 &     0 &     0 &     1 \\
      &     I-PB &    10 &     0 &     0 &     1 &     0 &     0 \\
      &     I-DY &     0 &     0 &     3 &     0 &     0 &     0 \\
B0.1L &      I-C &    10 &     0 &     0 &     0 &     0 &     0 \\
      &     E-PB &     1 &     0 &     0 &     1 &     0 &     0 \\
      &     E-DY &     0 &     0 &     0 &     0 &     0 &     0 \\
      &      E-C &     0 &     0 &     0 &     0 &     0 &     0 \\
\hline
      &      ISO &    32 &     0 &     1 &     1 &     0 &     0 \\
      &     I-PB &    32 &     0 &     0 &     0 &     0 &     0 \\
      &     I-DY &     0 &     0 &     1 &     0 &     0 &     0 \\
B0.3L &      I-C &    28 &     0 &     0 &     0 &     0 &     0 \\
      &     E-PB &     1 &     0 &     1 &     1 &     0 &     1 \\
      &     E-DY &     1 &     0 &     0 &     1 &     0 &     0 \\
      &      E-C &     0 &     0 &     1 &     0 &     0 &     0 \\
\hline
      &      ISO &    40 &     1 &     0 &     1 &     0 &     0 \\
      &     I-PB &    43 &     0 &     0 &     0 &     0 &     1 \\
      &     I-DY &     0 &     0 &     2 &     0 &     0 &     0 \\
B0.5L &      I-C &    39 &     0 &     0 &     0 &     0 &     0 \\
      &     E-PB &     0 &     0 &     0 &     0 &     0 &     1 \\
      &     E-DY &     0 &     0 &     2 &     0 &     0 &     0 \\
      &      E-C &     0 &     0 &     0 &     0 &     0 &     0 \\
\hline
      &      ISO &    64 &     0 &     2 &     1 &     0 &     3 \\
      &     I-PB &    55 &     0 &     0 &     0 &     0 &     1 \\
      &     I-DY &     5 &     0 &     1 &     0 &     0 &     0 \\
B0.7L &      I-C &    46 &     0 &     0 &     0 &     0 &     1 \\
      &     E-PB &     0 &     0 &     3 &     0 &     0 &     1 \\
      &     E-DY &     0 &     0 &     0 &     0 &     0 &     0 \\
      &      E-C &     0 &     0 &     1 &     0 &     0 &     0 \\
\hline
      &      ISO &    91 &     0 &     1 &     0 &     0 &     1 \\
      &     I-PB &    90 &     0 &     1 &     0 &     1 &     1 \\
      &     I-DY &     2 &     0 &     1 &     0 &     1 &     0 \\
B0.9L &      I-C &    74 &     0 &     0 &     0 &     0 &     1 \\
      &     E-PB &     5 &     0 &     2 &     1 &     0 &     1 \\
      &     E-DY &     0 &     0 &     0 &     0 &     0 &     0 \\
      &      E-C &     2 &     0 &     1 &     0 &     0 &     0 \\
\hline
      &      ISO &   201 &     0 &     1 &     1 &     0 &     0 \\
      &     I-PB &   182 &     0 &     1 &     3 &     0 &     0 \\
      &     I-DY &     1 &     0 &     2 &     0 &     0 &     0 \\
B1.0L &      I-C &   174 &     0 &     0 &     0 &     0 &     0 \\
      &     E-PB &     4 &     0 &     0 &     1 &     0 &     0 \\
      &     E-DY &     0 &     0 &     1 &     0 &     1 &     0 \\
      &      E-C &     3 &     0 &     0 &     0 &     0 &     0 \\
\hline
\hline
    \end{tabular}
    \label{tab:mergers}
\end{table}

The binary heating process is also the mechanism to produce GW sources, especially for the dynamically formed BBH mergers.
Thus, we also investigate how the GW mergers depend on $\fb$ based on the L-group models.
The numbers of mergers for different combination of WD, NS and BH binaries are summarized in Table~\ref{tab:mergers}.
For a self-consistent comparison, we only select mergers before 7~Gyr for all models. 
For each model, we also distinguish the origins of mergers by two ways, which are represented as the suffixes and prefixes in the type names, respectively.

\textit{Suffix}: we detect the mergers from primordial binaries (PB) and from dynamically formed binaries (DY). 
The former can be triggered either by the binary stellar evolution or by the stellar dynamics.
Thus, we also use the standalone \textsc{bse} code to directly evolve all primordial binaries from the initial conditions of each model and detect the GW mergers (ISO).
Some primordial binaries merged in both ISO and $N$-body models, we name it as common mergers (C).
In Table~\ref{tab:mergers}, we show the numbers of common mergers from $N$-body models.
But the component stellar types can be different for ISO and $N$-body models.
For example, the E-C in B0.3L has one BBH merger, but in the ISO group, it is a double-WD binary (BWD) merger.

\textit{Prefix}: meanwhile, mergers in the $N$-body models can be inside star clusters (the prefix "I") or already escape from the clusters (the prefix "E").

\subsection{The issue of the interface to \textsc{bse}}

Table~\ref{tab:mergers} shows that there is no BBH merger from primordial binaries in both ISO and $N$-body models (see the numbers of I-C and E-C in the BBH column).
We have investigated and found that this is due to a bug in the interface between \textsc{petar} and \textsc{bse}.
This bug appears when the main function (\textit{evolv2} in the source file) of \textsc{bse} is called multiple times.
\textsc{bse} is originally designed as a standalone code so that \textit{evolv2} is called only once to evolve a binary to a given time.
When the code is plugged in a $N$-body code, the \textit{evolv2} function needs to be called many times to couple the binary stellar evolution and the dynamical perturbation to the binary. 
We found that when \textit{evolv2} is called to evolve a binary in the Roche-overflow phase, a parameter is not properly initialized so that an artificial expansion of the orbit happens.
Such an expansion changes the following evolution and finally reduce the merger rate.
%\tanik{(AT: We may need to explain the bug in more detail. Because the BSE code is widely-used in the community, BSE users may be concerned by it. At least, we need to say 1) the cause of the bug, 2) the result of the bug, and 3) when the bug appears.)}
Unfortunately it is extremely computationally expensive to redo the $N$-body models (a few months), and thus, we can only redo the binary stellar evolution by using the fixed standalone \textsc{bse} code, similar to the ISO groups in Table~\ref{tab:mergers}.
The comparison between the old (ISO) and the bug-fixed versions are shown in Table~\ref{tab:bsefix}.
The bug-fixed version of \textsc{bse} results in much more BBH mergers, especially for the B1.0L model.
We expect that if the correct version is used in $N$-body models, the BBH mergers from primordial binaries would also increase.
Since we focus on the dynamically formed BBHs, although there is a systematical bias of BBH merger numbers, it does not change our major conclusion.
In the following analysis, we take into account the impact from this bug. 
 
\begin{table}
    \centering
    \caption{The comparison of the merger numbers by using the old standalone \textsc{bse} version (OLD) which is also used in the $N$-body models and the fixed standalone \textsc{bse} version (FIX). The table style is similar to Table~\ref{tab:mergers}.}
    \begin{tabular}{llcccccc}
\hline
Model &     Type &   BWD &   BNS &   BBH &  BWN &  BWB &  BNB \\
\hline
B0.1L &      OLD &    10 &     0 &     0 &     0 &     0 &     1 \\
      &      FIX &    12 &     0 &     7 &     1 &     0 &     1 \\
\hline
B0.3L &      OLD &    32 &     0 &     1 &     1 &     0 &     0 \\
      &      FIX &    30 &     0 &     3 &     0 &     0 &     0 \\
\hline
B0.5L &      OLD &    40 &     1 &     0 &     1 &     0 &     0 \\
      &      FIX &    47 &     0 &     4 &     0 &     0 &     2 \\
\hline
B0.7L &      OLD &    64 &     0 &     2 &     1 &     0 &     3 \\
      &      FIX &    67 &     0 &     4 &     1 &     0 &     1 \\
\hline
B0.9L &      OLD &    91 &     0 &     1 &     0 &     0 &     1 \\
      &      FIX &    89 &     0 &     3 &     1 &     0 &     1 \\
\hline
B1.0L &      OLD &   201 &     0 &     1 &     1 &     0 &     0 \\
      &      FIX &   193 &     1 &    21 &     2 &     0 &     0 \\
\hline
    \end{tabular}
    \label{tab:bsefix}
\end{table}

\subsection{Mergers via binary stellar evolution}

Table~\ref{tab:mergers} shows that most of the GW mergers are from BWDs, and no BWD merger is from dynamically formed binaries.
This is expected since BWD is not strongly affected by the stellar dynamics when BBHs dominate the binary heating as shown in Figure~\ref{fig:cbinrc}.
We expect that after the core collapse of light objects when most BHs have escaped, the dynamically formed BWDs will start to appear. 
Because BWDs formed from low-mass binaries, a larger $\fb$ results in a larger number of BWD mergers.

The numbers of BWD mergers are comparable in the ISO and in the $N$-body models and most of them occurred in both cases.
There are a few ISO BWD mergers that do not appear in the $N$-body models and vice versa. 
We check the history of the binary stellar evolution and dynamical encounters of these mergers, and find three reasons that cause the diverged evolution:

(1) The binary stellar evolution of the same binary is not fully reproducible when the calling frequency of the evolution function (\textit{evolv2}) is different. 
The different calling frequency can cause the numerical variation.
For example, the mass loss rate of the two components may be slightly different case by case.
Then, the difference grows up and the final result diverges.
Particularly, if the evolution track of one component in a binary is close to the boundary that can either evolve to a WD or a NS, the binary may become a BWD merger, a BWN merger or not a merger.  
The bug we found in the \textsc{bse} code is also related to the calling frequency and also contribute to the divergency.

(2) The asymmetric stellar wind and supernovae result in a natal kick when a WD, NS or BH forms.
In \textsc{bse}, the kick velocity is randomly sampled based on an assumption of the velocity distribution function. 
Such a random effect can significant change the final state of a binary, especially for the mergers including NSs or BHs.

(3) In star clusters, the dynamical interaction can also perturb the binary orbit and affect the final result. But this effect is weak for most BWD mergers.
It is mostly important for the BBH mergers. 
%However, due to the bug of \textsc{bse}, the BBH mergers from primordial binaries and binary stellar evolution channel are lack in the $N$-body models. 
%Thus, we cannot analyze this impact for these  BBH mergers.
%Thus, we cannot analyze this impact for BBH mergers.

\subsection{Mergers from dynamical formed binaries}

The number of DYN mergers (mainly from BBHs) are much less than that of the PB mergers and show a weak dependence on $\fb$.
As shown in Figure~\ref{fig:bhrhrc}, the evolution of $a$ for BBHs is similar for all models with primordial binaries.
Thus, although it may be affected by the low number statistics, it is reasonable to see a similar number of DYN mergers.

There are a few escaped mergers, where most are also PB mergers (see E-PB in Table~\ref{tab:mergers}).
In $N$-body models, these mergers are driven by a mix effect of binary stellar evolution and stellar dynamics. 
%The binary stellar evolution is more important for BWD mergers while stellar dynamics is more important for BBHs mergers.

\begin{figure*}
    \centering
    \includegraphics[width=\textwidth]{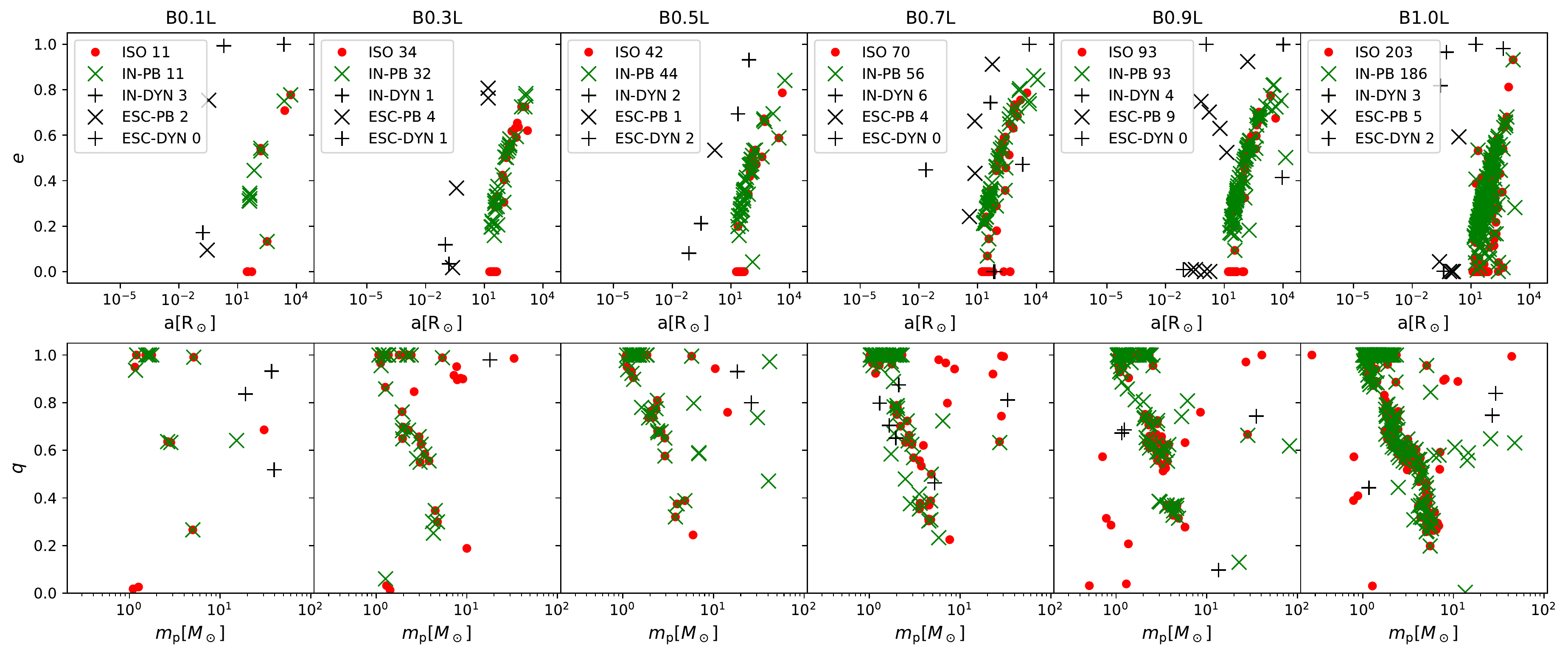}
    \caption{The orbital properties of the GW mergers for the L-group models excluding the B0L. 
    The upper panels show the semi-major axes and eccentricity ($e$) and the lower panels show the mass ratio ($q$) and the maximum mass of the two components in binaries ($m_{\mathrm{p}} [M_\odot]$). Colors and markers indicate the type of mergers: "ISO" represents the mergers before 7~Gyr drive by the isolated binary evolution performed by the standalone \textsc{bse} code; "IN" represents the mergers before 7~Gyr inside star clusters and "ESC" represents the escapers that can merge inside 12~Gyr.
    The suffix "-PB" represents the primordial binaries and "-DYN" represents the dynamically formed binaries.}
    \label{fig:gw}
\end{figure*}

In Figure~\ref{fig:gw}, we compare the initial (zero-age) orbital parameters of the GW mergers. 
The dynamical mergers show the initial orbital parameter when the binaries form.
The ISO and PB mergers have the $a$-eccentricity ($e$) distribution close to the boundary where the peri-center separation is close to the stellar radii of two components.
The escaped mergers from primordial binaries (E-PB) have very different $a$ and $e$ from the boundary, thus these mergers are probably driven by the dynamical interactions. 

Generally, we find that the low-mass binary (BWD) mergers do not show pronounced difference in the ISO and in the $N$-body models.
When massive BBHs exist, they dominate the binary heating process by which the dynamical impact on the low-mass binaries is strongly suppressed, as shown in Figure~\ref{fig:cbinrc}.
We expect that the dynamical channel to form low-mass GW mergers can only be important when  most BHs are kicked out from the host star clusters.

\section{Conclusions}
\label{sec:conclusion}

In this work, by using the high-performance $N$-body code, \textsc{petar}, we carry out a group of $N$-body simulations for the intermediate mass GC with initially $10^5$ stars, a half-mass radius of 2~pc and different fractions of primordial binaries ($\fb$ is up to 100 percent). 
The low-mass and massive primordial binaries are initialized differently.
In the L- and S-groups of models, the fraction of OB binaries is fixed to 100 percent while the fractions of low-mass binaries vary.
In the R-group for comparison, the fraction of binaries from all mass range changes.
In real star clusters, the evolution would be similar to the models with primordial binaries in the L- and S-groups, if most of OB stars were born in binaries.
We aim at understanding how primordial binaries affect the binary heating process, which influences the long-term dynamical evolution of the star cluster and the formation of GW mergers.

We find that massive primordial binaries composing OB stars have pronounced impact on the long-term structural evolution, such as $\rh$ and $\rc$ (Figure~\ref{fig:rhrc},\ref{fig:comps}), while the low-mass binaries do not.
The major reason is that when massive primordial binaries exist, more massive stars (BH progenitors) and BHs have escaped before 100 Myr via the few-body interaction between massive binaries, and thus, much less BHs were retained in the star cluster (Figure~\ref{fig:bhrhrc}, Table~\ref{tab:obesc}).
Because the BBHs dominate the binary heating process, the different numbers of BHs in between the models with and without primordial binaries result in different long-term structural evolution \cite[e.g.][]{Breen2013}.
Meanwhile, when BBHs exist, low-mass binaries have almost no chance to contribute to the binary heating, as shown in Figure~\ref{fig:cbinrc}.
Thus, the structural evolution of the star cluster is not sensitive to the fraction of low-mass binaries when BBHs dominate the binary heating.

To confirm this explanation, we also investigate how the stochastic effect from the random sampling of initial masses, positions and velocities of stars influences the evolution of $\rh$ and $\rc$.
The result (Figure~\ref{fig:comps}) shows that the scatter of the evolution due to the stochastic effect overlaps the difference caused by different $\fb$ (for low-mass binaries).

When the properties and fractions of massive binaries vary as in the R-group models, we observe an different evolution of $\rh$ and $\rc$ compared to that from the S- and the L-groups.
By comparing $\ncbh$ between the S- and the R-groups, we notice that the initial orbital properties of massive binaries have a stronger impact on $\ncbh$ than the fractions.

We also investigate how the stellar dynamics affects the GW mergers (Table~\ref{tab:mergers}) and compare the numbers of GW mergers from the $N$-body models and from the standalone \textsc{bse} code assuming all primordial binaries evolve in isolated environment (ISO).
Although the numbers of mergers are different in the ISO and in the $N$-body models, it is not caused by the stellar dynamics.
The major reason for the difference comes from the issue of numerical reproduciblity of the \textsc{bse} code and the random kick velocity after supernovae which change the binary orbits case by case.
Considering this variation, the number of BWD mergers are identical in the ISO and in the $N$-body models, while a few BBH mergers can be generated by the stellar dynamics.
The dependence on $\fb$ is weak. 
The BNS and BWD mergers from the dynamical channel become important only after most BHs are depleted and the core collapse of light objects (including NSs and WDs) occurs.

This study suggests that when BH subsystem exists, the long-term structural evolution of the star clusters is independent of the fraction of low-mass primordial binaries. 
Because the large binary fractions can significantly affect the computing performance of $N$-body simulations for massive GCs, our result provides a useful way to reduce such a difficulty, i.e. the $N$-body simulations only need to include massive primordial binaries, which is only a small fraction of the primordial binaries. 
On the other hand, the binary stellar evolution of low-mass binaries can be studied isolated by using the standalone binary stellar evolution codes. 
Such a way is valid before the core collapse of light objects (when most BHs are depleted).

\section*{Acknowledgements}
L.W. thanks the financial support from JSPS International Research Fellow (Graduate School of Science, The University of Tokyo). M.F. was supported by The University of Tokyo Excellent Young Researcher Program.
This work was supported by JSPS KAKENHI Grant Numbers 17H06360, 19H01933, and 19K03907, and MEXT as “Program for Promoting Researches on the Supercomputer Fugaku” (towards a unified view of the universe: from large scale structures to planets, revealing the formation history of the universe with large-scale simulations and astronomical big data).
Numerical computations were carried out on Cray XC50 at Center for Computational Astrophysics, National Astronomical Observatory of Japan and Fugaku at RIKEN center for computational science.

\section*{Data Availability}
The simulations underlying this article are performed on Cray XC50 and Fugaku.
The data on XC50 is transferred to the personal data server of the first author.
The data were generated by the software \textsc{petar}, which is available in GitHub, at https://github.com/lwang-astro/PeTar.
The simulation data will be shared via  private communication with a reasonable request.

%%%%%%%%%%%%%%%%%%%% REFERENCES %%%%%%%%%%%%%%%%%%

% The best way to enter references is to use BibTeX:

\bibliographystyle{mnras}
%\bibliography{example} % if your bibtex file is called example.bib

% Alternatively you could enter them by hand, like this:
% This method is tedious and prone to error if you have lots of references

%%%%%%%%%%%%%%%%%%%%%%%%%%%%%%%%%%%%%%%%%%%%%%%%%%

%%%%%%%%%%%%%%%%% APPENDICES %%%%%%%%%%%%%%%%%%%%%

%\appendix
%
%\section{Some extra material}
%
%If you want to present additional material which would interrupt the flow of the main %paper,
%it can be placed in an Appendix which appears after the list of references.

%%%%%%%%%%%%%%%%%%%%%%%%%%%%%%%%%%%%%%%%%%%%%%%%%%

% Don't change these lines
\bsp	% typesetting comment
\label{lastpage}
\end{document}